\documentclass[a4paper,11pt]{article}
\pdfoutput=1 % if your are submitting a pdflatex (i.e. if you have
\usepackage{jheppub} % for details on the use of the package, please
\usepackage[T1]{fontenc} % if needed

\usepackage[multiple]{footmisc}

\usepackage{graphicx}
\usepackage{amssymb, amsmath,mathtools,amsthm,mathrsfs}
\usepackage{hyperref}
\hypersetup{
    colorlinks=true,
    linkcolor=black,          % color of internal links (change box color with linkbordercolor)
    citecolor=black,        % color of links to bibliography
    filecolor=black,      % color of file links
    urlcolor=black          
}

\newcommand{\be}{\begin{equation}}
\newcommand{\ee}{\end{equation}}

\DeclareFontFamily{U}{MnSymbolC}{}
\DeclareSymbolFont{MnSyC}{U}{MnSymbolC}{m}{n}
\DeclareFontShape{U}{MnSymbolC}{m}{n}{
    <-6>  MnSymbolC5
   <6-7>  MnSymbolC6
   <7-8>  MnSymbolC7
   <8-9>  MnSymbolC8
   <9-10> MnSymbolC9
  <10-12> MnSymbolC10
  <12->   MnSymbolC12}{}
\DeclareMathSymbol{\intprod}{\mathbin}{MnSyC}{'270}

\def\tr{{\rm tr}}

\newcommand{\beq}{\begin{equation}}
\newcommand{\eeq}{\end{equation}}

\newcommand{\lieg}{\mathfrak{g}}

\newcommand{\poincare}{Poincar\'{e}}

\newcommand{\mink}{{\rm Mink}}

\title{\boldmath Kinematic space and the orbit method}

\author[1]{Robert F. Penna}
\affiliation[1]{Department of Physics, Columbia University\\
538 West 120th Street, New York, NY 10027} 
\emailAdd{rp2835@columbia.edu}

\author[1,2]{and Claire Zukowski}
\affiliation[2]{Institute for Theoretical Physics, University of Amsterdam\\
Science Park 904, 1098XH Amsterdam, The Netherlands} 
\emailAdd{c.e.zukowski@uva.nl}
\abstract{Kinematic space has been defined as the space of codimension-$2$ spacelike extremal surfaces in anti de Sitter (AdS$_{d+1}$) spacetime which, by the Ryu-Takayanagi proposal, compute the entanglement entropy of spheres in the boundary CFT$_d$. It has recently found many applications in holography. Coadjoint orbits are symplectic manifolds that are the classical analogues of a Lie group's unitary irreducible representations. We prove that kinematic space is a particular coadjoint orbit of the $d$-dimensional conformal group $SO(d,2)$. In addition, we show that the Crofton form on kinematic space associated to AdS$_3$, that was shown to compute the lengths of bulk curves, is equal to the standard Kirillov-Kostant symplectic form on the coadjoint orbit. Since kinematic space is K\"{a}hler in addition to symplectic, it can be quantized. The orbit method extends the kinematic space dictionary, which was originally motivated through connections to integral geometry, by directly translating geometrical properties of holographic auxiliary spaces into statements about the representation theory of the conformal group.

}

\begin{document} 
\maketitle
\flushbottom

\section{Introduction}

There has been significant progress over the past decade in understanding the way approximately local bulk fields are encoded in the CFT in holography~\cite{Hamilton:2005ju,Hamilton:2006az,Kabat:2011rz,Kabat:2017mun}. At the same time, there are obstructions to constructing diffeomorphism-invariant local operators in quantum field theory~\cite{Donnelly:2016rvo}. The intrinsic nonlocality that appears when gravity is coupled with quantum mechanics has led to interest in exploring non-local bulk observables that have interesting CFT interpretations.

One such program focuses on \emph{kinematic space}, an auxiliary space for holography that can be defined variously in terms of the set of spacelike bulk geodesics, codimension-$2$ spacelike bulk extremal surfaces or causal diamonds on the boundary (for a clarification on why these different choices lead to the same geometry for an AdS$_{d+1}$ bulk, see Section~\ref{sec:higherdimensions}). The original prescription for a static $3$-dimensional bulk was motivated by hole-ography, the program of computing the lengths of closed curves in the bulk from differential entropy~\cite{Balasubramanian:2013lsa,Czech:2014wka,Headrick:2014eia,Myers:2014jia}. Through connections to integral geometry, it was proposed that kinematic space is equipped with a symplectic form known as the Crofton form whose integral computes the lengths of bulk curves~\cite{Czech:2015qta}. 

Subsequently, certain bilocal operators in the CFT -- including the modular Hamiltonian for ball-shaped regions -- were seen to obey a simple wave equation on kinematic space~\cite{deBoer:2015kda,Czech:2016xec,deBoer:2016pqk}. Many additional results such as the HKLL formula for bulk reconstruction, a derivation of the linearized Einstein equation, the relation of conformal blocks to geodesic Witten diagrams, and even connections to the MERA tensor network also naturally emerged from this reorganization of holography~\cite{Czech:2016xec,Czech:2016tqr,Mosk:2016elb,Czech:2015kbp,Czech:2015xna}. 

Most of the kinematic space literature to date has focused on either pure AdS or locally AdS spaces such as the BTZ black hole, conical singularity and wormhole geometries~\cite{Czech:2015kbp, Czech:2015xna, Asplund:2016koz,Zhang:2016evx,Cresswell:2017mbk,Cresswell:2018mpj}. The one exception is defect geometries that are dual to boundary conformal field theories~\cite{Czech:2016nxc,Karch:2017fuh}. The extent to which the proposal is meaningful (or not) beyond highly symmetric and static examples remains an important question. 

Additionally, while there was an interesting connection made to the field of integral geometry, there has been little work done to understand the symplectic and K\"{a}hler structure of kinematic space. At the same time, the discovery of a natural Berry connection on kinematic space that is related to the lengths of bulk curves~\cite{Czech:2017zfq} should suggest some interpretation as a quantum phase space along the lines of~\cite{Ashtekar:1997ud}.

\begin{figure}
\centering
\includegraphics[width=0.65\textwidth]{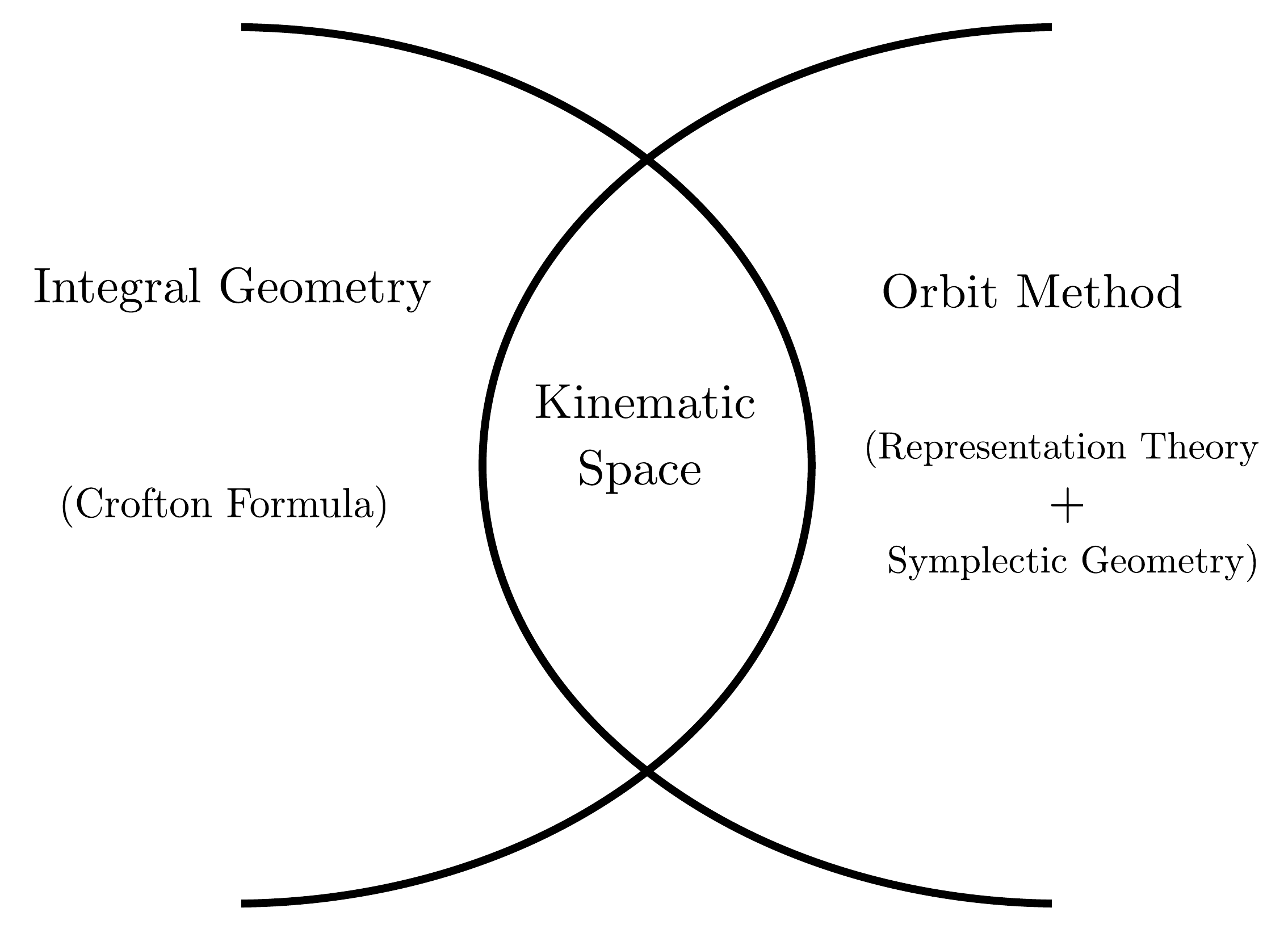}
\caption{Kinematic space has connections to two distinct branches of mathematics. The original motivation came from integral geometry. In this paper we point out a second connection to the orbit method, which lies at the crossover between representation theory and symplectic geometry.}
\label{fig:venndiagram}
\end{figure}

In this paper, we attempt to address both questions by proving that the kinematic space of AdS$_{d+1}$ is a coadjoint orbit of the $d$-dimensional conformal group. As we explain in Section~\ref{sec:orbits}, a coadjoint orbit is a symplectic manifold that is a homogeneous space of a given Lie group. The symplectic form associated to the coadjoint orbit is given by \eqref{eq:kirillov}. For certain types of orbits, in particular (but not exclusively) ones which additionally admit a K\"{a}hler structure, quantization by the orbit method equates the coadjoint orbit with a specific representation of the Lie group. The Atiyah-Bott formula \eqref{eq:atiyahbott} expresses characters of the associated representation as a sum over fixed points of vector fields on the coadjoint orbit. From the kinematic space perspective, this can be seen as an additional element of the kinematic space dictionary that converts geometrical statements about auxiliary spaces constructed from geodesics and other bulk surfaces into representation theory data.

{\bf Relation to Previous Work}: Part of our work focuses on the kinematic space associated to a time slice of AdS$_3$, where the full $SO(2,2)$ bulk symmetry is broken to the subgroup $SO(2,1)$. Coadjoint orbits of the double cover SL$(2,\mathbb{R})$ of $SO(2,1)$ have been studied in the past, see for example~\cite{Witten:1987ty, Vergne} for a review. The connection to kinematic space and the Crofton form is new. We also have not seen an explicit character computation in the literature using~\eqref{eq:atiyahbott} applied to the single-sheeted hyperboloid as we describe in Section~\ref{characters} (Witten~\cite{Witten:1987ty} discusses the analogous case for discrete series representations which correspond to the double-sheeted hyperboloid). In the higher dimensional case discussed in Section~\ref{sec:higherdimensions}, we consider coadjoint orbits of $SO(d,2)$ for $d>2$. As far as we know, these have not been studied in depth. It would be interesting to study these orbits and their quantization further in future work.

Finally, note that this is not the first time coadjoint orbits have appeared in holography; in the context of 3d gravity, for example, see~\cite{Maloney:2007ud, Oblak:2017ect, Barnich:2017jgw, Cotler:2018zff}, or in AdS$_2$/CFT$_1$ with connections to the SYK model see~\cite{Alekseev:1988ce, Mandal:2017thl, Stanford:2017thb}.  Coadjoint orbits have also been used to study the asymptotic symmetry groups of asymptotically flat spacetime \cite{Barnich:2014kra,Barnich:2015uva,Oblak:2015sea}. Our work differs from these, however, because we consider coadjoint orbits of the conformal group rather than the full asymptotic symmetry group. The resulting physical interpretation is also very different: we consider a phase space of geodesics on a fixed background, whereas coadjoint orbits of the full asymptotic symmetry group should be interpreted as a phase space of different \emph{metrics}.

{\bf Outline}: The paper is structured as follows. We begin in Section~\ref{sec:orbits} with a summary of the orbit method, including an illustration of these methods in the simple example of $SO(3)$. In Section~\ref{sec:symplectic}, we apply the orbit method to the case of a bulk AdS$_3$ spacetime where the kinematic space of geodesics on a slice is simply dS$_2$. We show that the Crofton form that computes bulk lengths matches the Kirillov-Kostant symplectic form on the coadjoint orbit, and discuss the K\"{a}hler structure. Section \ref{sec:higherdimensions} gives a partial generalization of these results to the $d$-dimensional case. We conclude by discussing some directions for future work, including implications for generalizing kinematic space beyond highly symmetric cases.  In Appendix~\ref{sec:timelike} we discuss the case of the space of timelike geodesics, which is equal to a different coadjoint orbit of the conformal group so that much of these techniques can be applied there as well.

\section{The orbit method}
\label{sec:orbits}

Let $G$ be a Lie group.  If a quantum system has $G$ symmetry, then its states lie in representations of $G$.  The analogue of unitary irreducible representations in the classical world are coadjoint orbits, which are a family of symplectic manifolds attached to any Lie group.  This section provides a brief introduction to the theory of coadjoint orbits and its relationship to representation theory.  For further details, see \cite{Vergne,Witten:1987ty,kirillov2004lectures, Oblak:2016eij}.  

Let $\lieg$ be the Lie algebra of $G$ and let $\lieg^*$ be the dual of the Lie algebra.  Then $\lieg^*$ is the space of linear maps, $\lieg\rightarrow k$, where in this paper $k$ can be $\mathbb{R}$ or $\mathbb{C}$.  Define a pairing, $\lieg\times \lieg^* \rightarrow k$, by
\beq\label{eq:pairing}
\langle \mu,\xi \rangle \equiv \mu(\xi)~, 
\eeq
where $\mu\in \lieg^*$ and $\xi\in \lieg$.  

Define a Poisson bracket on $\lieg^*$, called the Lie-Poisson bracket, by
\beq\label{eq:liepoisson}
\{F,G\}(\mu) \equiv \left\langle \mu , \left[\frac{\delta F}{\delta \mu},\frac{\delta G}{\delta \mu}\right] \right\rangle~.
\eeq
$F$ and $G$ are functions on $\lieg^*$ and their functional derivatives are considered as elements of $\lieg$.  The bracket $[\cdot,\cdot]$ is the Lie bracket on $\lieg$.  The Lie-Poisson bracket \eqref{eq:liepoisson} is clearly bilinear and antisymmetric, and it can be checked that it is a derivation and satisfies the Jacobi identity.  This makes $\lieg^*$ a Poisson manifold.   A Hamiltonian vector field, $X_H$, is defined in the usual way: it is a vector field on $\lieg^*$ for which there exists a function, $H$, such
\beq
X_H[G] = \{G,H\}~,
\eeq 
for all functions $G$.  Flows of Hamiltonian vector fields on $\lieg^*$ are classical dynamical systems.

The Lie-Poisson bracket defines a foliation of $\lieg^*$ into a family of symplectic leaves.  
Two points in $\lieg^*$ are said to be on the same symplectic leaf if there is a piecewise smooth curve in $\lieg^*$ joining the points such that each segment of the curve is a locally defined Hamiltonian vector field.  
The coadjoint orbits of $G$ can be succinctly described as the symplectic leaves of the Lie-Poisson bracket on $\lieg^*$.  
The restriction of the Lie-Poisson bracket \eqref{eq:liepoisson} to a coadjoint orbit is symplectic.  

The reason for the name coadjoint orbit is as follows.  The adjoint action of $G$ on $\lieg$ by conjugation is
\beq
g\cdot \xi = g\xi g^{-1}~,
\eeq
with an infinitesimal version
\beq
\xi\cdot \xi' = [\xi,\xi']~.
\eeq
This defines a map from $G$ to the automorphism group of $\mathfrak g$ called the \emph{adjoint representation}.

The coadjoint action of $G$ on $\lieg^*$ is defined using the pairing \eqref{eq:pairing},
\beq
\langle g\cdot \mu, \xi \rangle = \langle \mu,g^{-1} \cdot \xi \rangle~,
\eeq
for all $\xi\in \lieg$, with an infinitesimal version
\beq
\langle \xi\cdot \mu,\xi' \rangle = -\langle \mu, [\xi,\xi'] \rangle~,
\eeq
for all $\xi'\in \lieg$. This defines a map to the automorphism group of $\mathfrak g^*$ called the \emph{coadjoint representation}. It turns out that the orbits of the coadjoint action of $G$ on $\lieg^*$ coincide with the symplectic leaves of the Lie-Poisson bracket (see \cite{marsden2002introduction}). 

At a point, $\mu$, on a coadjoint orbit, we identify tangent vectors, $\tilde{\xi}$, with Lie algebra elements, $\xi\in \lieg$, using $\xi\cdot \mu = \tilde{\xi}$.  The Kirillov-Kostant symplectic form acts on tangent vectors as
\beq\label{eq:kirillov}
\omega_K(\tilde{\xi},\tilde{\xi'}) \equiv \langle \mu,[\xi,\xi'] \rangle~.
\eeq
It is the symplectic form induced by the restriction of the Lie-Poisson bracket to the orbit.  It plays a key role in all that follows.  

\subsection{Example: $SO(3)$}

\begin{figure}[t!]
	\centering
	\includegraphics[width=2.8in]{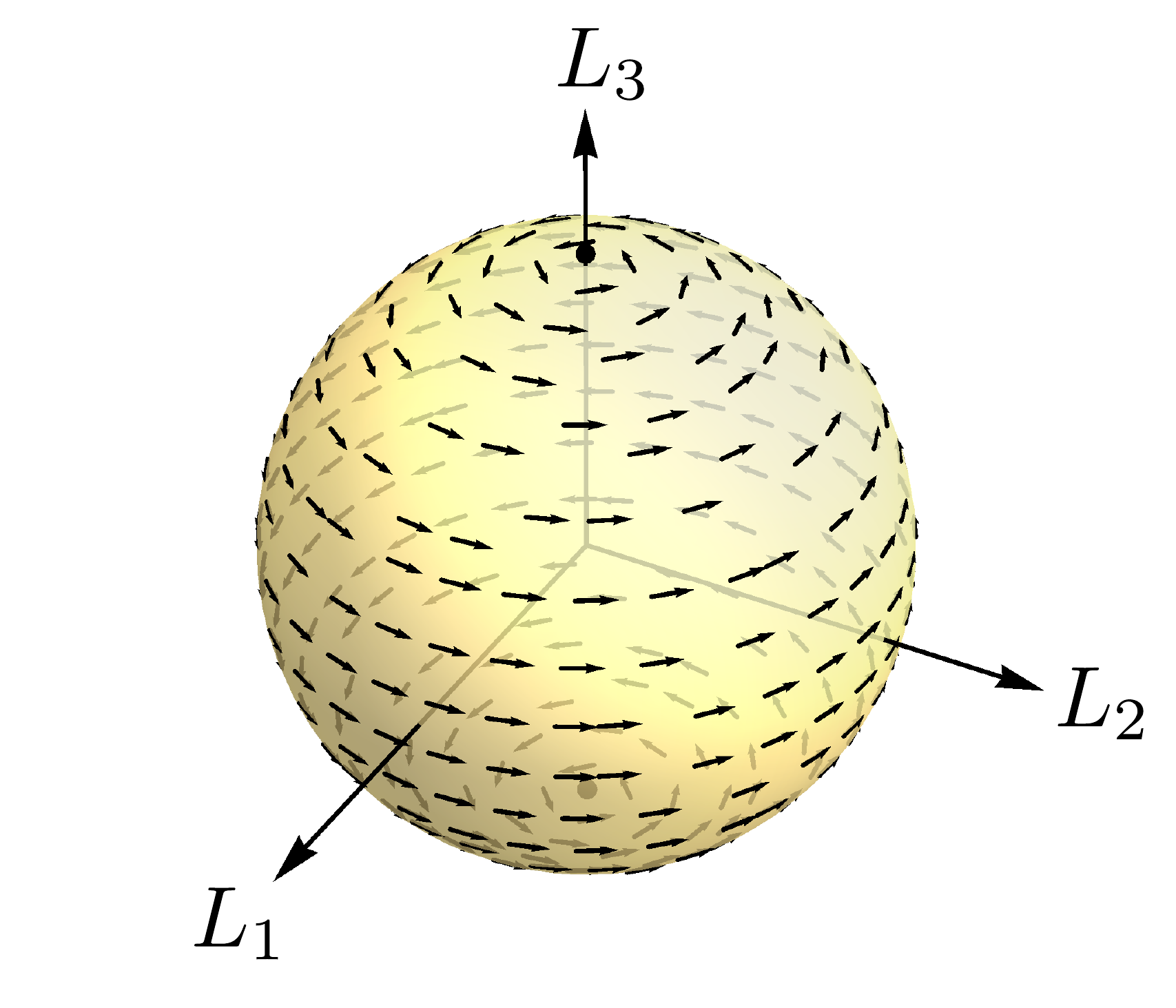}
	\caption{A coadjoint orbit of SO$(3)$ is a sphere. The irreducible unitary representations of $SO(3)$ come from quantizing the spheres with integral surface  area.  The vector field $\xi=\partial_\phi$ has two zeros, the north and south poles of the sphere.  The character, $\chi(q^{\hat{\xi}})$, is given by the Atiyah-Bott formula in terms of data at the fixed points.  This can be regarded as a path integral on the orbit computed using the method of stationary phase.}
	\label{spherevectors}
\end{figure}

A simple example incorporating all of these structures is the rigid body.   The configuration space of a rigid body is $G=SO(3)$.  Canonical phase space is the cotangent bundle, $T^*G\cong G\times \lieg^*$; this is ``orientation-angular momentum'' space.  The Euler equations for rigid body motion depend on the angular momentum variables only.  Quotienting out the orientation variables reduces the dynamics to Hamiltonian flow on $\lieg^*$.  In this process, the canonical Poisson bracket on $T^*G$ descends to the Lie-Poisson bracket on $\lieg^*$.

The Lie algebra, $\mathfrak{g} = \mathfrak{so}(3)$, is generated by $L_i$, $i=1,2,3$, which satisfy the commutation relations 
\be [L_i, L_j] = \epsilon_{ijk} L_k~.\ee 
There is an associated nondegenerate (because $\mathfrak{so}(3)$ is semisimple), $G$-invariant, positive definite bilinear Killing form on $\mathfrak{so}(3)$, which is defined\footnote{As $\mathfrak{so}(3)$ is a simple Lie algebra, there is a unique (up to a scalar multiple) invariant symmetric bilinear form.  So the trace can be taken in any irreducible representation.  For nonsimple algebras, the Killing form is defined by taking the trace in the adjoint representation.} on the basis elements as
\be \left<L_i,L_j\right> = -\frac{1}{2} \mbox{tr}(L_i L_j) = \delta_{ij}~.\ee  
We identify $\mathfrak{so}(3)\cong \mathfrak{so}(3)^*$ using the Killing form.  

The dual Lie algebra is $\lieg^*\cong \mathbb{R}^3$, and we can further identify $\mathfrak{so}(3)\cong \mathbb{R}^3$, in which case the Lie bracket is just the cross product. Each point represents an angular momentum vector, $\vec{J}=(J_1,J_2,J_3)$, for the rigid body.   The total angular momentum, $J^2$, is conserved in time.  So rigid body dynamics lies on two-spheres of the foliation $\mathbb{R}^3\cong S^2 \times \mathbb{R}^+$ of constant total angular momentum.  Likewise, the coadjoint action, $g\cdot J = gJ$, is just the ordinary action of the rotation group on $\mathbb{R}^3$ and the orbits are just the two-spheres of the foliation. It is important to note that the definition of coadjoint orbits does not depend on our choice of Hamiltonian: any Hamiltonian flow on $\lieg^*$ must lie on a coadjoint orbit.

From \eqref{eq:kirillov} we can compute the Kirillov-Kostant symplectic form. It is just the area element on the two-sphere divided by radius, $|\mu|$,
\be \omega_K = \frac{dA}{|\mu|} = |\mu|\sin \theta d\theta \wedge d\phi~,\ee
where $\theta$ and $\phi$ are the usual polar angles on the sphere. Indeed, the area element acts on tangent vectors as 
\beq
dA(\tilde{\xi},\tilde{\xi'}) = \hat{\mu} \cdot (\tilde{\xi} \times \tilde{\xi}')~,
\eeq
where $\hat{\mu}$ is the unit outward normal at the point $\mu$ on the sphere. Identify $\tilde{\xi} = \xi \times \mu$ and  $\tilde{\xi}' = \xi' \times \mu$.  Then
\beq\label{eq:proof}
dA(\tilde{\xi},\tilde{\xi'}) = \hat{\mu} \cdot [(\xi\times\mu) \times (\xi'\times\mu)]
	= |\mu| \mu \cdot (\xi\times \xi')
	= |\mu| \omega_K(\tilde{\xi},\tilde{\xi'})~.
\eeq
Later, when we turn to kinematic space, we will study coadjoint orbits of $SO(2,1)$.  Again we will find $\omega_K=dA/|\mu|$, where $dA$ is the  area form on the orbit.

\subsection{Representation theory}

Coadjoint orbits can be quantized using geometric quantization, giving unitary irreducible representations of $G$.  Not all orbits admit a quantization.  Those that do are called ``integral.''  The unitary irreducible representations of $SO(3) $ form an infinite discrete series, $\rho_\ell$, with $\ell=0,1,2,\dots$ and ${\rm dim}(\rho_\ell)=2\ell+1$.  These come from quantizing the two-spheres with integral surface area, where the surface area is measured with respect to  Liouville measure, $\omega_K/(2\pi)$.  The dimension of the representation is the surface area of the corresponding coadjoint orbit.

\begin{table}[t!]\label{tab:orbits}
\centering
\begin{tabular}{cc}
Representation Theory			&	Symplectic Geometry \\
\hline\\
irreducible unitary representations, $\rho$	&	integral coadjoint orbits, $\mathcal{O}$ \\\\
dim($\rho$)						&	area($\mathcal{O}$) \\\\
characters, $\chi_\rho$				&	path integrals, $\chi_{\mathcal{O}}$\\\\
Weyl character formula \cite{pressley1986loop}	&	Atyiah-Bott fixed point formula \eqref{eq:atiyahbott}
\end{tabular}
\caption{A summary of several aspects of the orbit method that we make use of in this paper.}
\end{table}

Given a representation, $\rho:G\rightarrow GL(V)$, of a Lie group on a vector space, one defines the character 
\beq\label{eq:chi}
\chi_\rho(g) \equiv \tr(\rho[g])~.
\eeq
Physically, characters are partition functions: let $g=e^{\vartheta\xi}$, with $\vartheta\in \mathbb{R}$, and think of  $\xi\in\lieg$ as the ``Hamiltonian''.  The orbit method gives a alternate method for computing characters: realize $\chi_\rho$ as a path integral on the coadjoint orbit corresponding to $\rho$.  It turns out the path integrals arising this way can be computed exactly using the method of stationary phase.  In particular, the Atiyah-Bott fixed point formula gives the character, $\chi_\rho(e^\xi)$, as a sum over zeros of $\xi$, where now $\xi$ is viewed as a vector field on the orbit.

Let us give the recipe in the case when the orbit, $\mathcal{O}_\rho$, is a complex manifold with holomorphic coordinates $z^1,\dots,z^k$.  Near each fixed point, expand in holomorphic coordinates as $\hat{\xi} \equiv -i\xi = n_k z_k\partial_{z_k}+\dots$.  Let $s$ be the number of negative $n_k$.  Define $q=e^{i\vartheta}$.  Then the Atiyah-Bott fixed point formula is 
\beq\label{eq:atiyahbott}
\chi_{\mathcal{O}_\rho}(q^{\hat{\xi}}) = \sum_{\text{zeros of $\xi$}}(-1)^s\frac{q^h}{1-q^{|n_k|}}~,
\eeq
where the Hamiltonian function, $h$, is defined in the usual way as
\beq
dh \equiv \omega_K(\xi,\cdot) = \iota_\xi \,\omega_K~.
\eeq

For $SO(3)$, we saw that the Kirillov-Kostant form is $\omega_K = dA/|\mu| = d\phi\wedge d\mathcal{Z}$, where $\mathcal{Z}=|\mu|\cos\theta$ and and $\theta$ and $\phi$ are the coordinates on the sphere.  The Hamiltonian function corresponding to the rotation $\xi=\partial_\phi$ is $h=\mathcal{Z}+{\rm const.}$ We convert to complex coordinates $z=\tan{(\theta/2)} e^{-i\phi}$ near the north pole, and $z=\cot{(\theta/2)}e^{i\phi}$ near the south pole. In terms of these the rotation becomes $\hat \xi = \pm (z\partial_z - \bar z \partial_{\bar z})$ (the minus sign corresponds to the north pole, while the plus sign corresponds to the south pole).

Now $\partial_\phi$ has zeroes at the north and south pole (Figure \ref{spherevectors}), where $h=\pm |\mu|+1/2$ (we have fixed the integration constant to its conventional value).  On the orbit with $|\mu|=\ell+1/2$, the Atiyah-Bott formula gives
\beq
\chi_{\ell}(q^{\hat{\partial_\phi}}) = -\frac{q^{|\mu|+1/2}-q^{-|\mu|+1/2}}{1-q}
	=\frac{q^{\ell+1/2}-q^{-(\ell+1/2)}}{q^{1/2}-q^{-1/2}}~.
\eeq
This can be checked directly from the definition \eqref{eq:chi}.  To see this, realize the representation $\rho_\ell$ as the $(2\ell+1)$-dimensional space of degree-$\ell$ spherical harmonics, $Y_{\ell m}$.  The rotation generator acts by $Y_{\ell m}\rightarrow e^{im\vartheta}Y_{\ell m}$.  So the character is $e^{-i\ell\vartheta}+\dots+e^{i\ell\vartheta}=(q^{\ell+1/2}-q^{-(\ell+1/2)})/(q^{1/2}-q^{-1/2})$, as above.

\section{Kinematic space is a coadjoint orbit}
\label{sec:symplectic}

In this section, we begin by introducing the kinematic space of a fixed time slice of AdS$_3$. Next we will identify this kinematic space with an adjoint orbit of $SO(2,1)$ and show that the symplectic structure on kinematic space is the same as the standard Kirillov-Kostant symplectic structure on the orbit. For now our definitions and results are restricted to this simple example, but keep in mind that we will return to discuss the higher dimensional generalization of kinematic space in Section~\ref{sec:higherdimensions}.

\subsection{Kinematic space for AdS$_3$}\label{kinematicspace}

Consider a CFT$_2$ in the vacuum on the cylinder, $\mathbb{R}\times S^1$. Each interval has an entanglement entropy, defined as $S_{\rm ent}(u,v)=-\tr(\rho_{uv}\log\rho_{uv})$, where $\rho_{uv}$ is the reduced density matrix on the interval.  Now identify the circle of the CFT$_2$ with the boundary of a fixed time slice of static AdS$_3$ with metric
\be ds^2 = -\left(\frac{r^2}{\ell^2}+1\right)dt^2 + \left(\frac{r^2}{\ell^2}+1\right)^{-1}dr^2 +r^2 d\phi^2~,\ee
where $\ell$ is the AdS radius. Intervals on the circle are in one-to-one correspondence with spacelike geodesics in the AdS$_3$ time slice with endpoints $u$ and $v$. By the Ryu-Takayanagi proposal, the entanglement entropy can be computed from the length of these geodesics and is given by
\beq
S_{\rm ent}(u,v) = \frac{c}{3}\log\frac{\Sigma}{\epsilon \pi}\sin\frac{\pi L}{\Sigma}~,\label{ee}
\eeq
where $\Sigma=2\pi R$ is the circumference of the circle, $L=R(v-u)$ is the length of the interval, $\epsilon$ is a UV cutoff, and $c=3\ell/2G$ is the Brown-Henneaux central charge.

We define kinematic space $\mathcal K$ in this setting as the space of all intervals $[u,v]\subset S^1$, or equivalently as the space of geodesics on the bulk time slice. This auxiliary space was originally motivated through connections with integral geometry~\cite{Czech:2015qta}. To see this, consider a curve $\gamma$ on the time slice in the bulk. Its length can be computed as an integral over kinematic space,
\be \frac{\rm{length}(\gamma)}{4G} = \frac{1}{4}\int_{\mathcal K} \omega_C(u,v) n_\gamma(u,v)~.\label{Croftonformula}\ee
Here $n_\gamma(u,v)$ is the number of times a given geodesic labeled by its endpoints $u,v$ intersects the curve $\gamma$, and $\omega_C(u,v)$ is a natural symplectic form on kinematic space known as the Crofton form. In terms of the entanglement entropy it is given by
\be \omega_C = \frac{\partial^2 S_{ent}(u,v)}{\partial u \partial v} du \wedge dv~.\label{eq:crofton}\ee
Using the explicit expression~\eqref{ee} for the entanglement entropy, this simplifies to
\beq
\omega_C = \frac{c}{12} \csc^2\left(\frac{v-u}{2}\right)du\wedge dv~.\label{eq:omegaC}
\eeq

The geodesics on the time slice (equivalently, their boundary intervals) obey a causal ordering: two geodesics (or intervals) are timelike related if one is contained in the other, they are null related if they share an endpoint, and they are spacelike related otherwise. This additionally defines a metric on kinematic space, 
\beq
ds^2 = \frac{c}{12} \csc^2 \left(\frac{v-u}{2}\right) dudv~,
\eeq
which is a metric for $2$-dimensional de Sitter space.

\subsection{Orbit interpretation}
\label{sec:so21}

The isometry group of AdS$_3$ is $SO(2,2)\cong SO(2,1) \times SO(2,1)$.  Fixing a constant time slice breaks the isometry group down to $SO(2,1)$.  In this section we will apply the orbit method to $SO(2,1)$.  The double cover of $SO(2,1)$ is $SL(2,\mathbb{R})$, and the orbit method description of the latter is well known (see e.g., \cite{Vergne,Witten:1987ty}).  The two cases are closely related.

\begin{figure}
\centering
\includegraphics[width=0.25\textwidth]{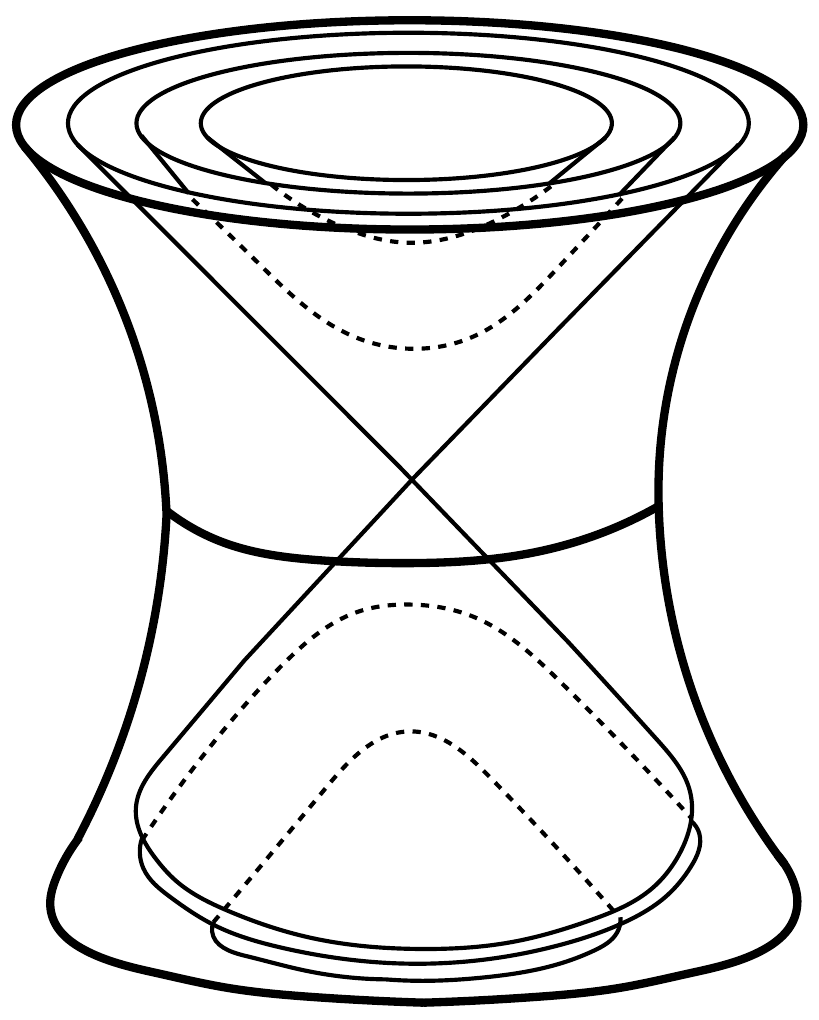}
\caption{Adjoint orbits of $SO(2,1)$.}
\label{fig:orbits}
\end{figure}

Like $SO(3)$, the group $SO(2,1)$ is semisimple,  so the Killing form is non-degenerate and we can use the Killing form to identify adjoint and coadjoint orbits.  Adjoint orbits are orbits of the group action on the Lie algebra, $\mathfrak{so}(2,1)$.   The Lie algebra is isomorphic to three dimensional Minkowski space. The isomorphism is the ``hat map,''
\beq
\hat{X} = 
\begin{pmatrix}
0	& X_2	& -X_1 \\
X_2	& 0		& -X_0\\
-X_1 & X_0	& 0\\
\end{pmatrix}
\leftrightarrow X = (X_0,X_1,X_2)~.
\eeq
Under this map, $\hat{X}Y=X\times_\eta Y$, where the $\mink_3$ cross product $X\times_\eta Y$ is the same as the usual Euclidean cross product, except the first component, $(X\times_\eta Y)^0$, differs by a minus sign.  It is straightforward to verify $[\hat{X},\hat{Y}]=(X\times_\eta Y)^{\widehat{}}\thinspace$, so the hat map is a Lie algebra isomorphism.

The adjoint action of $SO(2,1)$ on $\mink_3$ is given by Lorentz transformations.  Pick $X\in \mink_3$.  The adjoint orbit through $X$ is
\beq
\mathcal{O}_X \equiv \{g\cdot X | g\in SO(2,1)\} = \{Y\in \mink_3 | Y^2 = X^2\}~.
\eeq  
Evidently there are three cases to consider: positive, negative, and null $X^2$ (see Figure \ref{fig:orbits}).  Orbits with negative $X^2$ are double sheeted hyperboloids $\mathbb{H}^2\simeq SO(2,1)/U(1)$, orbits with positive $X^2$ are single sheeted hyperboloids dS$_2\simeq SO(2,1)/SO(1,1)$, and the orbit with $X^2=0$ is a cone through the origin.  Quantization of the double sheeted hyperboloids gives the discrete series representations of $SO(2,1)$.  Quantization of the single sheeted hyperboloids gives the principal series.  It is not known how to quantize the cone.  Also, the complementary series of representations of $SO(2,1)$ is missing; the orbit method does not seem to know about it (but see Kirillov \cite{kirillov2004lectures} for discussion of the possibility of an orbit interpretation of the complementary series).
So, the correspondence between orbits and irreducible representations is incomplete, at least as presently understood, but an orbit interpretation is available for the discrete and principal series of representations.

\subsection{Symplectic structure}

In Section~\ref{kinematicspace}, we recalled that the kinematic space for a time-slice in AdS$_3$ is a two-dimensional de Sitter space dS$_2$, and in Section \ref{sec:so21} we saw that this same space arose as one of the adjoint orbits of $SO(2,1)$.  We will now show that the Crofton form on kinematic space coincides with the Kirillov-Kostant symplectic form on the orbit.

\begin{figure}
\centering
\includegraphics[width=0.8\textwidth]{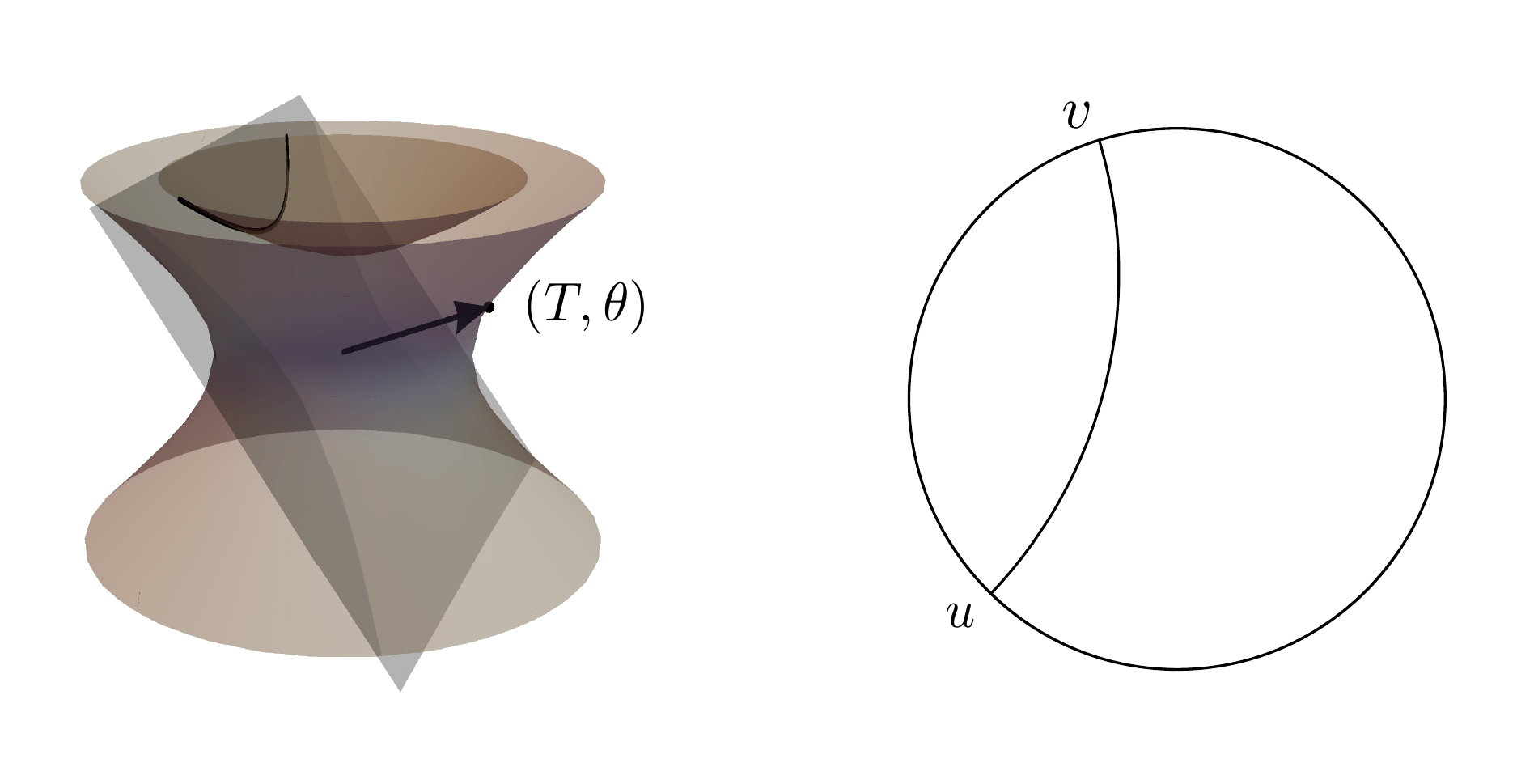}
\caption{\emph{Left:} Geodesics in the hyperbolic plane correspond to points, $(T,\theta)$, in the single sheeted hyperboloid.  \emph{Right:} The same geodesic projected onto the \poincare\ disk.}
\label{fig:pullback}
\end{figure}

The upper sheet of the double sheeted hyperboloid is a model for the hyperbolic plane and can be identified with a constant time slice of AdS$_3$.  As pointed out by  \cite{Czech:2015qta} in the context of kinematic space, there is a simple map between geodesics in the hyperbolic plane and points on the the single sheeted hyperboloid.  To define this map, note that a geodesic in the hyperbolic plane defines a plane through the origin of Mink$_3$ (see Figure  \ref{fig:pullback}).  The normal to the plane fixes a point in the single sheeted hyperboloid. 

It is helpful to introduce new coordinates on kinematic space, $u=\theta-\alpha$ and $v=\theta+\alpha$.  Note that $\theta$ is the polar angle of the midpoint of the geodesic and $\alpha$ is the half-angle subtended by the geodesic on the boundary.  In these coordinates, the Crofton form Eq.~\eqref{eq:omegaC} is
\beq\label{eq:crofton2}
\omega_C = \frac{c}{12}\csc^2\alpha \thinspace d\alpha \wedge d\theta~.
\eeq

Now let us compare with the Kirillov-Kostant form on the orbit. The dS$_2$ hyperboloid in $\mink_3$ is given by the embedding
\be -(X^0)^2 + (X^1)^2 + (X^2)^2 = \ell^2~,\ee
where $\ell$ is the de Sitter radius.

Introduce coordinates
\begin{align}
X^0 &= \ell \cot T~,\notag\\
X^1 &= \ell \cos \theta' \csc T~,\notag\\
X^2 &= \ell \sin \theta' \csc T~.
\end{align}
The induced metric is
\beq
ds^2 = \ell^2 \csc^2 T(-dT^2+d\theta'^2)~.
\eeq

From \eqref{eq:kirillov} we can compute the Kirillov-Kostant symplectic form. It is given by
\beq
\omega_K = \ell \csc^2 T dT \wedge d\theta'~.\label{eq:omegaK}
\eeq
Just as for SO$(3)$ orbits, this is is equal to $\omega_K = dA/|X| = dA/\ell$, where $dA$ is the area form. The proof goes the same as \eqref{eq:proof}, only now using the Mink$_3$ cross product introduced at the beginning of this section. Note that this is almost the same as \eqref{eq:crofton2} (up to an overall numerical factor).  We need to show $T=\alpha$ and $\theta=\theta'+{\rm const.}$ under the map described by Figure \ref{fig:pullback}.  The equality $\theta=\theta'+{\rm const.}$ is obvious.  

To show $T=\alpha$, we give a concrete construction of the map between kinematic space and the orbit.  Fix a point $\vec{X}$ on the hyperboloid.
Introduce vectors $\vec{X}'$ and $\vec{X}''$ which are orthogonal to $\vec{X}$ and each other (as Mink$_3$ vectors):
\begin{align}
\vec{X}' &= \frac{\ell/\sqrt{2}}{\sqrt{\cos{2T}+\cos{2\theta}}} \left(
2\cos\theta\cot T~,
\cos 2T\csc T+\cos 2\theta\csc T~,
2\cos\theta\sin\theta\csc T\right)~,\\
\vec{X}''  &= \frac{\ell}{\sqrt{\cos^2T-\sin^2\theta}}(\sin\theta,0,\cos T)~.
\end{align}
Note $\vec{X}^2=\vec{X}''^2=\ell^2$ and $\vec{X}'^2=-\ell^2$.

Now fix a point $(T,\theta)$ on the single sheeted hyperboloid and consider the plane spanned by $\vec{X}'$ and $\vec{X}''$.  The intersection of this plane with the hyperbolic plane is a geodesic, $\gamma(\tau)$.  The geodesic can be parametrized as
\beq
\vec{\gamma}(\tau) = (\gamma^1,\gamma^2,\gamma^3) =  \cosh\tau \vec{X}' + \sinh\tau \vec{X}''~.
\eeq
Projecting onto the \poincare\ disk gives
\beq
\vec{\gamma}_D(\tau) = \left(\frac{\ell \gamma^2}{\ell +\gamma^1},\frac{\ell \gamma^3}{\ell+\gamma^1}\right)~.
\eeq
The endpoints of the geodesic are
\beq
\vec{\gamma}_\pm \equiv \lim_{\tau\rightarrow \pm\infty} \vec{\gamma}_D(\tau) \equiv (\gamma_\pm^1,\gamma_\pm^2) = \ell \left(\cos{(T\pm \theta)}, \sin{(T\pm \theta)}\right)~.
\eeq
The angles are $\tan u=\gamma_-^2/\gamma_-^1$ and $\tan v=\gamma_+^2/\gamma_+^1$.  A straightforward but tedious calculation gives $\alpha = (v-u)/2 = T$, as desired.  This shows that the Crofton form and the Kirillov-Kostant form coincide (up to an overall numerical constant).

\subsection{K\"{a}hler structure}

In expression \eqref{eq:crofton} for the Crofton form,
\begin{equation*}
\omega_C  = \frac{\partial S_{\rm ent}(u,v)}{\partial u \partial v} du \wedge dv~,
\end{equation*}
the entanglement entropy, $S_{\rm ent}$, appears to play the role of a K\"{a}hler potential for the symplectic form.  The orbit interpretation of kinematic space sheds some light on this connection.  Coadjoint orbits are usually K\"{a}hler manifolds or cotangent bundles (these are essentially the only two cases for which it is understood how to quantize the orbit and get representations).  So the orbit interpretation of kinematic space makes it less surprising that the Crofton form comes from a  K\"{a}hler potential.  Strictly speaking the orbit is a Lorentzian K\"{a}hler manifold \cite{chen2011pseudo}, because the metric is not positive definite.  Let us make the interpretation of $S_{\rm ent}$ as a K\"ahler potential precise.

First recall that a K\"{a}hler structure is a special kind of complex structure.  A complex structure essentially means that our manifold locally looks like $\mathbb{C}^n$.  This means first of all that at each point, $p$, there is a linear map, $J_p$, on the tangent space such that $J_p^2=-1$.  This map should vary smoothly as a function of $p$; viewed as a $(1,1)$-tensor field, the components of $J_p$ are smooth functions.  For $\mathbb{C}$ itself, $J_p$ corresponds to multiplication by $i$.  The $+i$ and $-i$ eigenvectors of $J_p$ are called holomorphic and antiholomorphic vectors, respectively, and $J_p$ is called an almost complex structure.  To qualify as a complex structure, there must exist a coordinate induced basis for the holomorphic and antiholomorphic vectors.

A K\"{a}hler manifold has a complex structure, but it also has a metric and a symplectic form, and all three structures are compatible in a natural sense.  There are many equivalent ways of formulating the precise definition.  Here is one: first define a (pseudo)-Hermitian manifold to be a complex manifold with a (pseudo-)Riemannian metric, $g$, such that $g(JX,JY)=g(X,Y)$ for all tangent vectors, $X,Y$.   A (pseudo)-Hermitian manifold is (pseudo)-K\"{a}hler if the fundamental 2-form, $g(X,JY)$, is closed.

Now consider kinematic space, viewed as the single sheeted hyperboloid, $\mathcal{O}_\ell$.  Consider the linear map, $J:T\mathcal{O}_\ell\otimes \mathbb{C} \rightarrow T\mathcal{O}_\ell\otimes \mathbb{C}$, of the complexified tangent bundle, defined by
\begin{align}
\frac{\partial}{\partial \theta} &\rightarrow i \frac{\partial}{\partial \alpha}~, \\
i\frac{\partial}{\partial \alpha} &\rightarrow -\frac{\partial}{\partial \theta}~,
\end{align}
where the coordinates $(\alpha,\theta)$ are as above.  $J$ sends $\partial_u \rightarrow -i \partial_u$ and $\partial_v \rightarrow i \partial_v$.  In other words, $\partial_v$ and $\partial_u$ are holomorphic and antiholomorphic vectors with respect to the complex structure, $J$.  

The metric on the hyperboloid is
\beq
ds^2 = \ell^2 \csc^2 \left(\frac{v-u}{2}\right) dudv~.
\eeq
Clearly $g(JX,JY)=g(X,Y)$ for all $X,Y\in T\mathcal{O}_\ell\otimes \mathbb{C}$, so the orbit is a Lorentzian Hermitian manifold.   Furthermore, the fundamental 2-form, $g(X,JY)$, is the Kirillov-Kostant form (up to an overall constant) and therefore closed.  This makes the orbit a Lorentzian K\"{a}hler manifold.
Let $\partial$ and $\bar{\partial}$ be the Dolbeault operators corresponding to $u$ and $v$.  For example, acting on functions, $\partial f(u,v) = f_{,u} du$.  The Crofton form becomes simply
\beq\label{eq:kahler}
\omega_C = \partial \bar{\partial} S_{\rm ent}~,
\eeq
and $S_{\rm ent}$ is the K\"ahler potential.

\subsection{$SO(2,1)$ characters}\label{characters}

So far everything has been classical.  While the orbit method gives us a way to quantize kinematic space, it is not immediately clear whether this is physical, in the sense that the characters one computes using the orbit method correspond to physical partition functions. In this paper we will treat quantization as a tool that adds to the kinematic space dictionary. We leave a more complete exploration of the physicality of quantization to future work.

Now we will show how to compute characters of $SO(2,1)$ principal series representations using the orbit method. For background on the  representation theory of $SL(2,\mathbb{R})$, see \cite{Bargmann, HC}. The principal series consists of functions, $f(z)$, on the Poincar\'{e} disc which transform under the group action as
\begin{align} 
(\rho_s(g^{-1}) f)(z) &= |c z + d|^{-1+i s} f\left(\frac{az+b}{cz+d}\right)~,\\
(\rho_s(g^{-1}) f)(z) &= \mbox{sgn}(cz+d)|c z + d|^{-1+i s} f\left(\frac{az+b}{cz+d}\right)~,
\end{align}
where $s\in \mathbb{R}$ and $\left(\begin{array}{cc} a& b\\ c & d\end{array}\right)\in \mbox{SO}(2,1)$. The principal series corresponds to the single sheeted hyperboloid, $\mathcal{O}_\ell$ which we saw was a model for kinematic space.  It turns out that all of the $\mathcal{O}_\ell$ can be quantized, there is no further ``integrality'' condition on the orbit.

The first thing to note about the character of a representation \eqref{eq:chi} is that it is a class function: it only depends on the conjugacy class of $g$.  This follows immediately from the conjugation invariance of the trace.  Up to orientation, elements of $SO(2,1)$ fall into two conjugacy classes: rotations and Lorentz boosts.  So it will suffice to evaluate the character for a standard rotation and a standard boost.  The action of a rotation on the single sheeted hyperboloid has no fixed points.  So the Atiyah-Bott formula gives $\chi(g)=0$ when $g$ is a rotation.  And indeed this can be verified directly using the definition \eqref{eq:chi} and an explicit construction of the principal series representations.  

\begin{figure}[t!]
	\centering
	$\vcenter{\hbox{\includegraphics[width=1.75in]{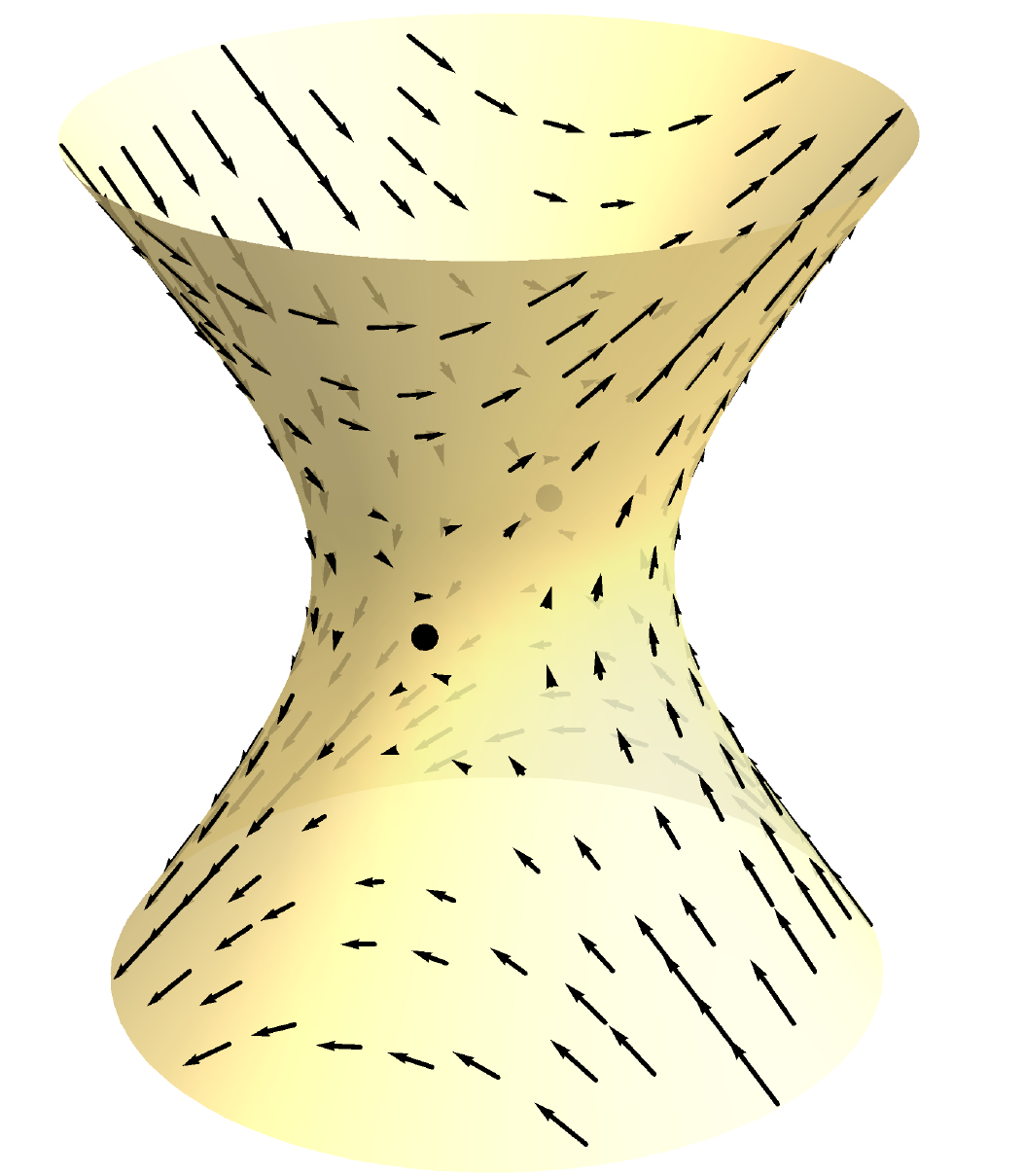}}}$
	\hspace{0.2in}
	$\vcenter{\hbox{\includegraphics[width=3.25in]{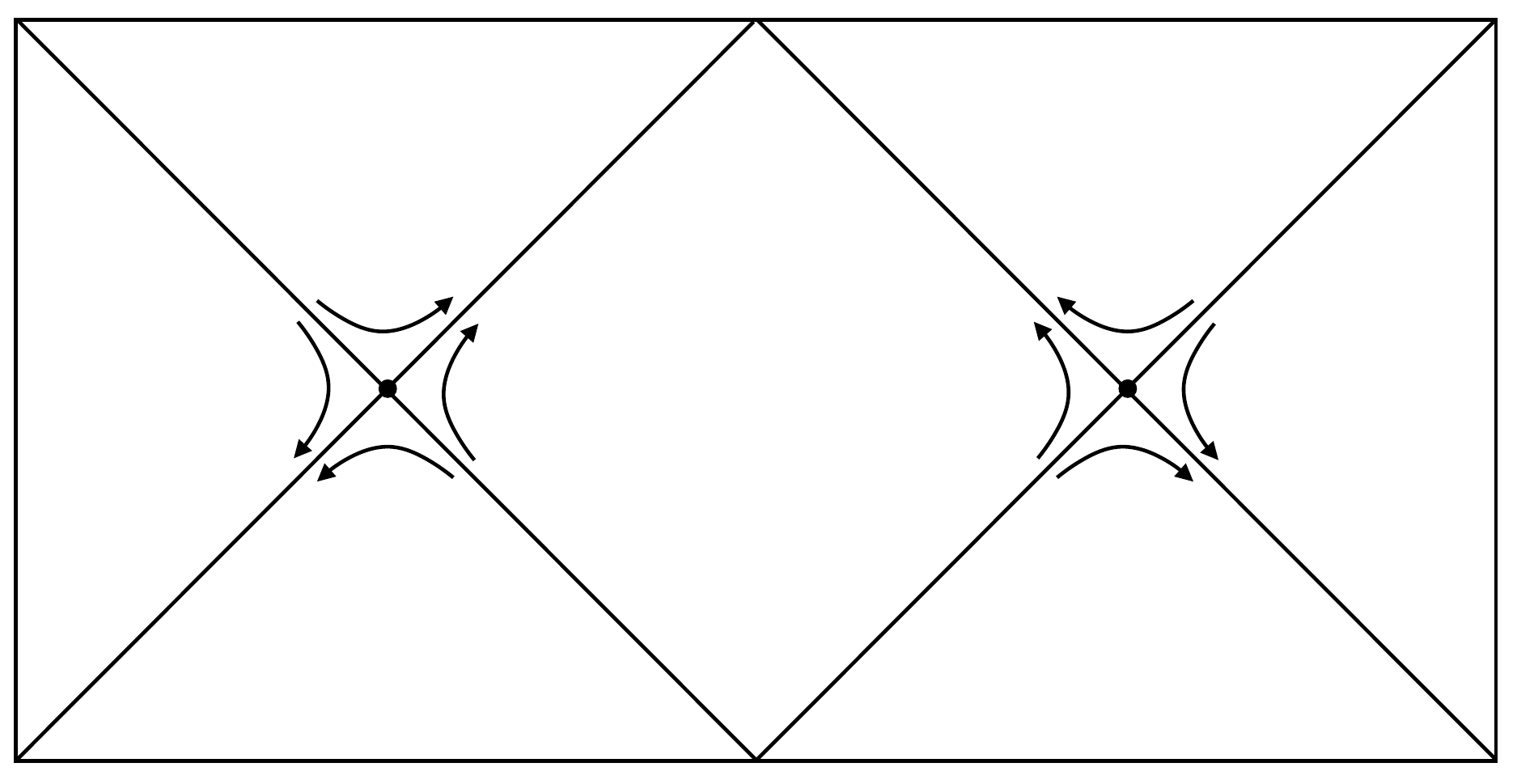}}}$
	\caption{Kinematic space is a particular coadjoint orbit of $SO(2,1)$ given by the two-dimensional de Sitter hyperboloid, depicted here on the embedding diagram (left) and Penrose diagram (right) for dS$_2$ along with the boost Eq.~\eqref{eq:boost}. Characters of the principal series representations can be computed geometrically from sums over the two fixed points of this boost using the Atiyah-Bott formula. }
	\label{boost}
\end{figure}

The more interesting case is when $g$ is a Lorentz boost.  Consider the action of the $\mink_3$ Lorentz boost
\beq\label{eq:boost}
X^0 \frac{\partial}{\partial X^1} + X^1 \frac{\partial}{\partial X^0}
\eeq
on the hyperboloid $-(X^0)^2+(X^1)^2+(X^2)^2=\ell^2$.   There are two fixed points: $(X^0,X^1,X^2)=(0,0,\pm \ell)$.  Introduce
static coordinates
\begin{align}
X^0 &= \sqrt{\ell^2-r^2}\sinh{\eta}~,\\
X^1 &= \sqrt{\ell^2-r^2}\cosh{\eta}~,\\
X^2 &= r~.
\end{align}
Near the fixed points, set $r=\pm \ell(1-R^2/2\ell^2+\dots)$, and so $X^0=R\sinh\eta, X^1=R\cosh\eta,$ and $X^2=\pm\ell$.  In these coordinates, the induced metric on the hyperboloid near the fixed points is just that of Rindler space:
\beq
ds^2 = -R^2 d\eta^2 + dR^2 + (\dots)
\eeq 
and the boost \eqref{eq:boost} is Rindler time translation, $\partial_\eta$.  Under a Wick rotation to Euclidean space this is an ordinary rotation with respect to an imaginary angle, $\eta=i \phi$ (see Figure \ref{boost}).

In static de Sitter coordinates, $dA=\ell d\eta\wedge dr$, and so the Kirillov-Kostant symplectic form is 
\beq\label{eq:omegadS}
\omega_K = \frac{dA}{\ell}= d\eta \wedge dr~.
\eeq

The Hamiltonian, $h$, corresponding to $i\partial_\eta$ is defined by
\beq
dh= \omega_K(i\partial_\eta,\cdot)~.
\eeq 
Plugging in \eqref{eq:omegadS} gives $h=\pm i\ell+1/2$ at the fixed points (an integration constant has been fixed for convenience).

We can convert from embedding coordinates to complex coordinates defined by
\be z' = \frac{X^1+i X^2}{\sqrt{\ell^2 + (X^0)^2}} e^{X^0/\ell}~.\ee
This maps the infinite past $X^0=-\infty$ of the hyperboloid to the origin and the waist $X^0=0$ to the circle $|z'|=1$. The two fixed points are at $z' = \pm i$. Taking $z = z' \mp i$ near the two fixed points and expanding in $R$, we find $\hat \xi = -i \partial_\eta = \bar z \partial_z - z \partial_{\bar z}+...$

We now have everything we need to compute the character for a standard boost\footnote{To get the signs and factors of $i$ right, it helps to think of this as a character for the rotation $\partial_\phi = i\partial_\eta$.} with rapidity $\zeta >0$.  Let $q=e^\zeta$.  The result is
\beq
\chi_\ell(q^{\partial_\eta}) 
	= -\frac{q^{i\ell+1/2} + q^{-i\ell+1/2}}{1-q}
	= \frac{q^{i\ell}+q^{-i\ell}}{q^{1/2}-q^{-1/2}}~.
\eeq
These are precisely the characters of the $SO(2,1)$ principal series \cite{Vergne}.  The characters blow up as $q\rightarrow 1$ because the representations are infinite dimensional.  This is reflected in the infinite area of the hyperboloids.

To match notation with Vergne \cite{Vergne}, identify $\ell=s/2$ (the factor of 2 appears because our definition of the symplectic form differs from Vergne's by a factor of 2). Furthermore, recall the Lie group isomorphism $SL(2,\mathbb{R})/\mathbb{Z}_2 \cong SO(2,1)$ and identify the Lie algebra, $\mathfrak{sl}(2,\mathbb{R})$, with $\mink_3$ via the map
\beq
\tilde{X}=\begin{pmatrix}
X_2	& X_1+X_0	\\
X_1-X_0	& -X_2		\\
\end{pmatrix}
\leftrightarrow (X_0,X_1,X_2)~.
\eeq
Now consider the adjoint action, $\tilde{X}\rightarrow g_\lambda\tilde{X}g_\lambda^{-1}$,  of 
\beq\label{eq:sl2boost}
g_\lambda=\begin{pmatrix}
e^\lambda	& 0 \\	
0	& e^{-\lambda} \\
\end{pmatrix}
\in SL(2,\mathbb{R})
\eeq
on $\mathfrak{sl}(2,\mathbb{R})$.  This sends
\begin{align}
X_0 &\rightarrow X_0 \cosh 2\lambda + X_1 \sinh 2\lambda~,	\notag\\
X_1 &\rightarrow X_1 \cosh 2\lambda + X_0 \sinh 2\lambda~,	\notag\\
X_2 &\rightarrow X_2~.
\end{align}
In other words, $g_\lambda$ acts as a Lorentz boost with rapidity $\zeta=-2\lambda$.  Trading parameters relates our notation and Vergne's.

\section{Higher dimensions}
\label{sec:higherdimensions}

In $d+1$-dimensional anti-de Sitter space, $AdS_{d+1}$, there are multiple equivalent definitions for kinematic space, $\Gamma_{2d}$. From the perspective of entanglement entropy, it makes sense to define $\Gamma_{2d}$ as the space of codimension-$2$ extremal surfaces or, equivalently, their boundary $(d-2)$-spheres. The areas of these extremal surfaces compute the entanglement entropy of boundary spheres by the Ryu-Takayanagi proposal. This is also the same as the space of causal diamonds on the boundary, which was studied in~\cite{deBoer:2016pqk}. The reason for this is that every causal diamond can be generated from the past and future development of pair of timelike related points, whose intersection is a $(d-2)$-sphere. It is also equally suitable to continue defining kinematic space as the space of spacelike geodesics, but in higher dimensions. See~\cite{Czech:2016xec} for a comparison. 

Kinematic space in arbitrary dimensions is a coset space. To review this, first consider the space of bulk codimension-$2$ extremal surfaces. From the embedding space perspective, bulk codimension-$2$ extremal surfaces result from the intersection of the AdS$_{d+1}$ hyperboloid in $\mathbb{R}^{2,d}$ with $d$-planes with one spacelike and one timelike normal passing through the origin. Asymptotically, this intersection yields the corresponding boundary $(d-2)$-spheres. The spheres are preserved by $SO(d-1,1)$ transformations acting on the $d$-plane as well as $SO(1,1)$ transformations orthogonal to it. Any sphere can be reached from another by a conformal transformation, modulo those that leave the spheres fixed. 

Next consider the space of spacelike geodesics. These are left invariant by $SO(1,1)$ translations along the geodesics, as well as $SO(d-1,1)$ boosts around the geodesics. The geodesics map between one another under any $SO(d,2)$ transformation modulo those that leave a geodesic fixed.

The resulting coset space in either case is kinematic space,
\be \Gamma_{2d} = \frac{SO(d,2)}{SO(d-1,1)\times SO(1,1)}~.\label{cosetspace}\ee 
It has a metric given by~\cite{Czech:2016xec,deBoer:2016pqk}
\be ds^2 = \frac{4 L^2}{(x-y)^2} \left(-\eta_{\mu\nu} + \frac{2(x_\mu-y_\mu)(x_\nu-y_\nu)}{(x-y)^2}\right) dx^\mu dy^\nu~, \label{higherdmetric}\ee
where $L$ is the radius of curvature and $x^\mu$ and $y^\mu$ describe either the timelike related pairs of points defining the tips of the causal diamond, or the pairs of spacelike boundary points that define a bulk geodesic. Notice that the signature is $(d,d)$, so this space is always even dimensional as is necessary for a symplectic structure.\footnote{Of course, even dimension does not guarantee a symplectic structure; the even-dimensional spheres, $S^{2n}$, with $2n>2$, are not symplectic.}

We will show that the kinematic space for AdS$_{d+1}$ is a coadjoint orbit of $SO(d,2)$. It is promising that it takes the form of~\eqref{cosetspace}, since the coadjoint orbits of a Lie group, $G$, are homogenous spaces of the form $G/K$, where $K$ is a subgroup of $G$.  When $G$ is compact and connected\footnote{Proposition 5.3 of \cite{kirillov2004lectures}.}, coadjoint orbits correspond precisely to those subgroups $K\subset G$ containing the maximal torus of $G$.  Since $SO(d,2)$ is not compact, it is not so obvious that $\Gamma_{2d}$ is a coadjoint orbit.  To check this, we need to find a coadjoint vector whose stabilizer is $SO(d-1,1)\times SO(1,1)$. 

First, note that  $\mathfrak{so}(d,2)$ has a nondegenerate invariant bilinear form, $\tr(XY)$ (where $X,Y\in\mathfrak{so}(d,2)$), so linear functions on $\mathfrak{g}$ may be identified with elements of $\mathfrak{g}$ and adjoint and coadjoint vectors can be identified.  So our task is to find an adjoint vector whose stabilizer is $SO(d-1,1)\times SO(1,1)$.  We will work infinitesimally and exhibit an adjoint vector whose stabilizer is $\mathfrak{so}(d-1,1)\times \mathfrak{so}(1,1)$.

To begin, let
\be\label{eq:gform}
g\equiv{\rm diag}(\underbrace{-1,\dots,-1}_{p},\underbrace{+1,\dots,+1}_{q})~.
\ee
The group $O(p,q)$ is the set of $n\times n$ matrices $A$ satisfying
\be
A^T g = g A^{-1}~,
\ee
where $n=p+q$.  So the Lie algebra $\mathfrak{o}(p,q)$ is the set of $n\times n$ matrices $X$ satisfying
\be
X^T g  = -gX~.
\ee
$X_{ij}$ is antisymmetric for $0<i,j\leq p$ and $p<i,j\leq n$, and symmetric on all other indices.  A representative element of $\mathfrak{o}(2,2)$ is
\be
\begin{pmatrix}
	0	&	-1	&	1	&	1	\\
	1	&	0	&	1	&	1	\\
	1	&	1	&	0	&	1	\\
	1	&	1	&	-1	&	0	
\end{pmatrix}~.
\ee

Introduce a basis for $\mathfrak{o}(p,q)$ as follows.  Let $\delta$ be the $n\times n$ Kronecker delta.  Define
\be
(M^{\rho\sigma})^{\mu\nu} = \delta^{\rho\mu} \delta^{\sigma\nu} - \delta^{\rho\nu} \delta^{\sigma\mu}~.
\ee
The $M^{\rho\sigma}$ are antisymmetric $n\times n$ matrices.  Let ${(M^{\rho\sigma})^\mu}_\nu = (M^{\rho\sigma})^{\mu\gamma}g_{\gamma\nu}$, with $g_{\gamma\nu}$ defined by \eqref{eq:gform}.

The ${(M^{\rho\sigma})^\mu}_\nu$ form a basis for $\mathfrak{o}(p,q)$.  For example, for $\mathfrak{o}(2,2)$, the basis is 
\begin{align}\label{eq:Mbasis}
({{M^{12}})^\mu}_\nu&=\begin{pmatrix}
	0	&	-1	&	0	&	0	\\
	+1	&	0	&	0	&	0	\\
	0	&	0	&	0	&	0	\\
	0	&	0	&	0	&	0	
\end{pmatrix}
\quad
({{M^{13}})^\mu}_\nu=\begin{pmatrix}
	0	&	0	&	+1	&	0	\\
	0	&	0	&	0	&	0	\\
	+1	&	0	&	0	&	0	\\
	0	&	0	&	0	&	0	
\end{pmatrix}
\quad
({{M^{14}})^\mu}_\nu=\begin{pmatrix}
	0	&	0	&	0	&	+1	\\
	0	&	0	&	0	&	0	\\
	0	&	0	&	0	&	0	\\
	+1	&	0	&	0	&	0	
\end{pmatrix}\notag\\
%---
({{M^{23}})^\mu}_\nu&=\begin{pmatrix}
	0	&	0	&	0	&	0	\\
	0	&	0	&	+1	&	0	\\
	0	&	+1	&	0	&	0	\\
	0	&	0	&	0	&	0	
\end{pmatrix}
\quad
({{M^{24}})^\mu}_\nu=\begin{pmatrix}
	0	&	0	&	0	&	0	\\
	0	&	0	&	0	&	+1	\\
	0	&	0	&	0	&	0	\\
	0	&	+1	&	0	&	0	
\end{pmatrix}
\quad
({{M^{34}})^\mu}_\nu=\begin{pmatrix}
	0	&	0	&	0	&	0	\\
	0	&	0	&	0	&	0	\\
	0	&	0	&	0	&	+1	\\
	0	&	0	&	-1	&	0	
\end{pmatrix}~.
\end{align}
Note $M^{\rho\sigma}=-M^{\sigma\rho}$, so there are six independent matrices in this example.  

Now $\mathfrak{so}(p,q)\cong \mathfrak{o}(p,q)$, so we have an explicit basis for $\mathfrak{so}(p,q)$.  Fix a basis element, $M^{\rho\sigma}$, and consider the adjoint orbit passing through it:
\beq\label{eq:orbitM}
\mathcal{O}_{M^{\rho\sigma}} \equiv \{g\cdot M^{\rho\sigma}|g\in SO(p,q)\}~.
\eeq
Our goal is to find an adjoint orbit that realizes kinematic space \eqref{cosetspace}. The orbits are quotient manifolds, $SO(p,q)/H$, where $H$ is the stabilizer group,
\beq
H\equiv{\rm Stab}(M^{\rho\sigma})\equiv\{g\in SO(p,q)|g\cdot M^{\rho\sigma}=M^{\rho\sigma}\}~.
\eeq
The Lie algebra of the stabilizer is
\be
\mathfrak{h}\equiv {\rm stab}(M^{\rho\sigma}) \equiv \{X\in \mathfrak{so}(p,q) \thinspace |\thinspace [X,M^{\rho\sigma}] = 0 \}~.
\ee
Clearly $M^{\rho\sigma}\in {\rm stab}(M^{\rho\sigma})$.  This generates a one-dimensional subgroup of the stabilizer.  We also have $M^{\rho'\sigma'}\in {\rm stab}(M^{\rho\sigma})$ whenever $\rho',\sigma',\rho$ and $\sigma$ are all distinct.  This generates an $(n-2)(n-3)/2$ dimensional subgroup of the stabilizer.  The stabilizer is the product of these two subgroups.

For example, the stabilizer of $M^{23}\in \mathfrak{so}(2,2)$ is
\beq
{\rm stab}(M^{23})={\rm span}\{M^{23},M^{14}\}=\mathfrak{so}(1,1)\times \mathfrak{so}(1,1)~.
\eeq
So $\mathcal{O}_{M^{23}}\cong SO(2,2)/(SO(1,1)\times SO(1,1))$ realizes kinematic space in two dimensions.

Similarly, kinematic space in three dimensions can be realized as an adjoint orbit of $\mathfrak{so}(3,2)$.  The stabilizer of $M^{34}\in \mathfrak{so}(3,2)$ is 
\beq
{\rm stab}(M^{34})={\rm span}\{M^{34},M^{12},M^{15},M^{25}\}=\mathfrak{so}(2,1)\times \mathfrak{so}(1,1)~.
\eeq
So $\mathcal{O}_{M^{34}}\cong SO(3,2)/(SO(2,1)\times SO(1,1))$ realizes kinematic space in three dimensions.  

In general, kinematic space in $d$ dimensions can be realized as the adjoint orbit of $SO(d,2)$ passing through $M^{d,d+1}$.
It was remarked in~\cite{deBoer:2016pqk} that the metric~\eqref{higherdmetric} on kinematic space in $d$ dimensions takes the K\"{a}hler-like form
\be ds^2 = \frac{\partial^2 V}{\partial x^\mu \partial y^\nu} dx^\mu dy^\nu \ee
with 
\be V = 2 L^2 \log{\left[-(x-y)^2\right]}~.\ee
We expect this to be related to the Kirillov-Kostant form in the adjoint orbit description but leave an exploration of this relationship for the future.

As discussed further in the Appendix, it is possible to define an analogue of kinematic space using timelike geodesics instead of spacelike geodesics in $AdS_{d+1}$.  This space is isomorphic to $SO(d,2)/(SO(2)\times SO(d))$ and it is not hard to check that it too can be realized as an adjoint orbit of $SO(d,2)$.  It is the orbit passing through $M^{d+1,d+2}$.

\section{Discussion}
\label{sec:discussion}

We have shown that the `kinematic space' of geodesics or extremal surfaces in AdS$_{d+1}$ has a natural symplectic structure due to the fact it is a coadjoint orbit of the $d$-dimensional conformal group. The existence of an additional K\"{a}hler structure ensures that it can be quantized, so through the orbit method kinematic space additionally maps to a representation of the conformal group. (In the case of AdS$_3$, the corresponding representation is a principal series representation.) Such tools add to the kinematic space dictionary by translating aspects of the representation theory of the conformal group into geometrical statements about holographic auxiliary spaces; for instance, we explained how characters can be computed from kinematic space using the Atiyah-Bott formula, \eqref{eq:atiyahbott}. 

Kinematic space has not been understood beyond simple examples such as pure AdS, the BTZ black hole, conical singularities and certain defect geometries. For a bulk geometry with less symmetry, one can always similarly define an auxiliary space of codimension-$2$ extremal surfaces or spacelike geodesics, however it is not clear that it would have useful properties. From the integral geometry perspective, one might hope there is a generalization of the Crofton form~\eqref{eq:crofton} that reconstructs bulk lengths in general spacetimes from an auxiliary space of geodesics or surfaces.   One motivation for this work was to either identify a structure that could be generalized to extend the Crofton formula~\eqref{Croftonformula} to less symmetric and non-static geometries, or to rule out such a generalization. 

\raggedbottom

From the point of view of coadjoint orbits, which carry a symplectic form~\eqref{eq:kirillov} that we have shown matches the Crofton form of~\cite{Czech:2015qta}, the relevant input is the asymptotic conformal symmetry. Beyond pure AdS and closely related spacetimes this would not be sensitive to details deep inside the bulk where the symmetry is broken. One might imagine that for special cases where some but not all of the bulk symmetry is broken it would be possible to instead consider coadjoint orbits of the broken isometry group, which would admit a candidate Crofton symplectic form given by \eqref{eq:kirillov}.  However, in situations with less symmetry, the information encoded in spaces of surfaces becomes more complicated while the coadjoint orbits of the remaining symmetry group contain less information.  So it is hard to see how this could work.  For example, consider the case where all but a bulk $U(1)$ symmetry is broken. The relevant coadjoint orbit is just a point. On the other hand, the lengths of curves would depend non-trivially on their radial location. This suggests moving away from viewing kinematic space as a tool for bulk reconstruction in arbitrary spacetimes. Instead, it may be more appropriate to interpret kinematic space as a phase space for holography defined in terms of the asymptotic conformal symmetry, which geometrizes the representation theory of the conformal group.

A more fruitful avenue for future work might be to consider the complementary approach of Alekseev-Shatashvilli path integral quantization of coadjoint orbits in the kinematic space context. In $3$d gravity the path integral quantization of the coadjoint orbit Diff($S^1$)/$PSL(2,\mathbb{R})$ of the Virasoro group~\cite{Alekseev:1988ce, Alekseev:2018ful} gives the Polyakov action that describes a theory of boundary gravitons in AdS$_3$~\cite{Cotler:2018zff, Barnich:2017jgw} (see also~\cite{Coussaert:1995zp, Henneaux:1999ib}). The Schwarzian model for the emergent gravitational mode in the SYK model dual to AdS$_2$ Jackiw-Teitelboim gravity can be seen as being embedded in this theory. From the kinematic space perspective, such techniques might help interpret some recent results connecting the equations of motion on kinematic space to Jackiw-Teitelboim gravity~\cite{Callebaut:2018xfu,Callebaut:2018nlq}. Additionally, in the Virasoro case the path integral approach has been used to compute Virasoro blocks as the two point function of bilocal operators invariant under a local $PSL(2,\mathbb{R})$ symmetry~\cite{Cotler:2018zff}. It would be interesting to connect this to the kinematic space story of OPE blocks, contributions to the OPE from a single conformal family which were seen to propagate as fields on kinematic space~\cite{Czech:2016xec}.

One especially intriguing aspect of the Crofton (equivalently Kirillov-Kostant) symplectic form associated to the AdS$_3$ kinematic space was its connection to entanglement entropy, which entered as a K\"{a}hler potential. Given that the orbit method applies equally well to the other coadjoint orbits of the conformal group, such as the space of timelike geodesics as well as spaces of other non-Ryu-Takayanagi bulk surfaces in AdS$_{d+1}$ that are stabilized by appropriate subsets of the AdS isometries, it would be interesting to understand if the K\"{a}hler potentials in these cases also have quantum information theoretic interpretations.\\

\noindent{\bf Acknowledgments}: It is a pleasure to thank Jan de Boer, Alejandra Castro, Bartlomiej Czech, Ben Freivogel, Kurt Hinterbichler, Lampros Lamprou, Blagoje Oblak, Charles Rabideau, Philippe Sabella-Garnier and James Sully for helpful discussions. CZ is grateful to the workshop ``From Kinematic Space to Bootstrap: Modern Techniques for CFT and AdS'' and the Mainz Institute for Theoretical Physics (MITP) for its hospitality and partial support during the completion of some of this work. CZ is supported by NASA ATP grant NNX16AB27G and ERC Consolidator Grant QUANTIVIOL.  RFP is supported by a Prize Postdoctoral Fellowship in the Natural Sciences at Columbia University and by Simons Foundation Award Number 555117.

\appendix

\section{Timelike geodesics}
\label{sec:timelike}
The symplectic structure of timelike geodesics in AdS$_3$ has been considered in~\cite{Gibbons:1999rb} and the relevant coadjoint orbits and their quantization have been discussed in many references including~\cite{Witten:1987ty}. Here, we provide a summary and make an explicit connection between the two.

AdS$_{d+1}$ can be obtained as the hyperbola in $\mathbb{R}^{2,d}$ obtained by the embedding
\be -(X^0)^2 - (X^1)^2 + (X^2)^2 +... + (X^{d-2})^2 = -\mathcal \ell^2~,\ee
where $\ell$ is the AdS radius. The metric is induced from the usual one on $\mathbb{R}^{2,d}$.

The timelike geodesics in AdS$_{d+1}$ can be obtained as the intersection of the set of totally timelike two-planes passing through the origin of $\mathbb{R}^{2,d}$. Such planes are invariant under $SO(2)$ timelike rotations along the plane as well as the $SO(d)$ orthogonal rotations. Thus, the space of timelike geodesics is equal to the Grassmannian $SO(d,2)$/$SO(2)\times SO(d)$.

Let us specialize to the case of AdS$_3$. Then the space of timelike geodesics factorizes into two copies of $SO(2,1)$/$SO(2)$, which can be identified as the double-sheeted hyperboloid. Recall that this is one of the other orbits of $SO(2,1)$ that was obtained in Section~\ref{sec:so21}.

By the orbit method the hyperboloid maps to a discrete series representation of $SO(2,1)$. These representations consist of bounded holomorphic functions $f(z)$ on the Poincar\'{e} disk which transform under the group action as
\be (\rho_n(g^{-1}) f)(z) = (c z + d)^{-(n+1)} f\left(\frac{az+b}{cz+d}\right)~, \indent (n\geq1)~,\label{discrete}\ee
where $n\in \mathbb{Z}$ and $\left(\begin{array}{cc}a & b\\ c & d\end{array}\right) \in SO(2,1)$.
For these representations, the integer $n$ is discrete so that the group acts in a single-valued way, and unitarity of the representation is guaranteed by restricting to $n\geq 1$.

The hyperboloid $\mathbb H_2$ in $\mathbb{R}^{1,2}$ can be obtained by the embedding 
\be -(X^0)^2 + (X^1)^2 + (X^2)^2 = -\ell^2~,\ee
where $\ell$ is the radius of the hyperboloid. The metric is induced from the usual one on $\mathbb{R}^{1,2}$. Choose embedding coordinates given by
\be
\vec{X}=(X^0,X^1,X^2) = \ell (\cosh{r}, \sinh{r}\cos\theta, \sinh{r}\sin{\theta})~
\ee
where $r\in (0,\infty)$ and $\theta \in (0,2\pi)$. We can further map from embedding space to coordinates $z = x+iy, \bar z = x-iy$ on the Poincar\'{e} disk $|z|<1$ by taking
\be
x = \frac{\ell X^1}{\ell + X^0}~, \indent y = \frac{\ell X^2}{\ell + X^0}~.
\ee

Let $\tilde{\eta},\tilde{\chi}$ be tangent vectors to this orbit at a point $p$. Relate $\tilde{\eta},\tilde{\chi}$ to adjoint vectors $\eta, \chi \in \mathfrak{so}(2,1)$ using the coadjoint action: 
$\tilde{\eta} = \eta \cdot p$ and $\tilde{\chi} =\chi\cdot p$.  Then the orbit symplectic form~\eqref{eq:kirillov} is
\be\label{eq:omegahyperbolicorbit}
\omega(\tilde{\eta},\tilde{\chi}) = \langle p,[\eta,\chi]\rangle~.
\ee
In intrinsic coordinates, this becomes
\be \omega=\ell \sinh r \, dr \wedge d\theta~.\label{hyperbolicomega}\ee

Let us compute the character corresponding to an SO$(2)$ rotation of the hyperboloid around the $X^0$-axis using the Atiyah-Bott formula. The rotation is generated by the vector field
\be X^1 \frac{\partial}{\partial X^2} - X^2 \frac{\partial}{\partial X^1}~.\ee

The Hamiltonian vector field $\partial_\theta$ defines a Hamiltonian function determined from Eq.~\eqref{hyperbolicomega} by
\be
dh = \iota_{\partial_\theta} dX^0 \wedge d\theta \nonumber = dX^0~.
\ee
Integrating and setting the constant to zero gives $h=X^0$. We have a single fixed point at $X^0=0$ on the bottom of the hyperboloid (or top if we consider the lower branch), where the vector field in complex coordinates takes the form $\hat \xi = -i \partial_\theta = z \partial_z - \bar z \partial_{\bar z}$. In~\eqref{eq:atiyahbott} $s=0$, $n_k=1$, $q=e^{i\theta}$ and the Hamiltonian at the fixed point is just $h=\ell$. Taking $2\ell = n+1$, we obtain
\be
\chi_n (q^{\partial_\theta}) = \frac{q^{(n+1)/2}}{1-q}~.\label{discretechar}
\ee

Now let us compare by computing the character directly from the discrete series transformation~\eqref{discrete}. In terms of the complex coordinate $z$, the rotation is $z\rightarrow e^{i\theta} z$. Representing it as an element $g\in SO(2,1)$, it corresponds to the choice $a=e^{i\theta/2}, d=e^{-i\theta/2}, b=c=0$. Under this rotation, the holomorphic functions transform as $f(z) \rightarrow q^{(n+1)/2} f(q z)$ with $q=e^{i\theta}$. A basis of holomorphic functions is given by the polynomial functions $1, z, z^2, ...$ Thus the character is given by the sum $\sum_{k=0}^\infty q^{(n+1)/2+k}$, which reproduces \eqref{discretechar}. Similar to the case of $SO(3)$ and unlike the single-sheeted hyperboloid, only orbits that satisfy a discretization condition on the radius of the hyperboloid, $2\ell= n+1$, can be quantized.

\bibliographystyle{utcaps}
\bibliography{KSCoadjoint}

\end{document}